\documentclass[aps,prl,reprint,superscriptaddress,10pt,longbibliography]{revtex4-1}

\usepackage{graphicx}
\usepackage{amsmath}
\usepackage{amssymb}
\usepackage{amsfonts}
\usepackage{bm}
\usepackage{bbm}
\usepackage[mathcal]{euscript}

\usepackage{multirow}

\usepackage{color}

\newcommand{\etal}{\textit{et al}.}

\begin{document}
\title{Growth mechanism and origin of high $sp^3$ content in tetrahedral amorphous carbon}

\author{Miguel~A. Caro}
\email{mcaroba@gmail.com}
\affiliation{Department of Electrical Engineering and Automation,
Aalto University, Espoo, Finland}
\affiliation{Department of Applied Physics, Aalto University, Espoo, Finland}
\author{Volker L. Deringer}
\affiliation{Engineering Laboratory, University of Cambridge,
Trumpington Street, Cambridge CB2 1PZ, United Kingdom}
\affiliation{Department of Chemistry, University of Cambridge,
Lensfield Road, Cambridge CB2 1EW, United Kingdom}
\author{Jari Koskinen}
\affiliation{Department of Chemistry and Materials Science,
Aalto University, Espoo, Finland}
\author{Tomi Laurila}
\affiliation{Department of Electrical Engineering and Automation,
Aalto University, Espoo, Finland}
\author{G\'abor Cs\'anyi}
\affiliation{Engineering Laboratory, University of Cambridge,
Trumpington Street, Cambridge CB2 1PZ, United Kingdom}

\newcommand{\eq}[1]{Eq.~(\ref{#1})}
\newcommand{\fig}[1]{Fig.~\ref{#1}}

\begin{abstract}
We study the deposition of tetrahedral amorphous carbon (ta-C) films from molecular
dynamics simulations based on a machine-learned interatomic potential trained from density-functional theory data.
For the first time, the high $sp^3$ fractions in excess of 85\% observed experimentally have been reproduced
by means of computational simulation and the deposition energy-dependence of the film's
characteristics is also accurately described. High confidence in the potential and
direct access to the atomic interactions allow us to infer
the microscopic growth mechanism in this material. While the widespread view is
that ta-C grows by ``subplantation'', we show that the so-called ``peening'' model
is actually the dominant mechanism responsible for the high $sp^3$ content. We show that pressure waves lead to bond
rearrangement away from the impact site of the incident ion, and high $sp^3$ fractions arise from a delicate
balance of transitions between 3- and 4-fold coordinated carbon atoms. These results open the door
for a microscopic understanding of carbon nanostructure formation with an unprecedented level of predictive power.
\end{abstract}

\date{\today}

\maketitle

Amorphous carbons (a-C) are a class of materials with important applications as coatings. Of special
interest are high-density forms of a-C which exhibit a high fraction of $sp^3$-bonded carbon atoms, known as tetrahedral a-C (ta-C)
or diamond-like carbon (DLC) because their mechanical properties are similar to those of diamond. Emerging applications
of a-C are as precursors in the synthesis of other forms of nanostructured carbons~\cite{sainio_2016b,suarez_2012}
and as a substrate
platform for biocompatible electrochemical devices~\cite{laurila_2017}. Significant efforts are being made to develop
carbon-based devices designed for biological sensing, which could be implantable in the human body, and will
be at the heart of the next technological revolution, where seamless integration between human tissue and
microelectronics will enable real-time health monitoring and countless other applications~\cite{tiwari_2015,arriaga_2016,laurila_2017}.

Together with its widespread technological and industrial use, a-C has also been the subject of significant academic interest,
in particular by the computational
modeling community. The high degree of bonding flexibility exhibited by carbon, which can exist in $sp^3$,
$sp^2$ and $sp$ environments or ``hybridizations'', is behind its ability to form numerous compounds which make the sheer complexity of
life possible. This flexibility is also responsible for the large degree of microscopic variability found in a-C,
where diverse and disordered atomic motifs can coexist, each in its own metastable configuration.
This makes simulations of a-C a long-standing challenge for any computational model based on interatomic potentials.
Early molecular dynamics (MD)
studies focused on optimizing and parameterizing simple classical potentials for a-C~\cite{tersoff_1988}, but also seminal
\textit{ab initio} MD (AIMD) simulations of a-C were conducted when the field was still in its infancy~\cite{galli_1989,kaukonen_1992}.
A constant struggle for computational models, since early on and until today, has been to recreate and understand the formation
process which leads to the high $sp^3$ fractions observed for ta-C, which can be in excess of 85\%.
Experimentally, ta-C is commonly grown by deposition of energetic ions onto a substrate. The fraction of $sp^3$
carbon increases monotonically with the beam energy up to approximately 60~eV--100~eV (depending on the method)~\cite{robertson_2002},
where it peaks at around 90\%. At
higher energies, the amount of $sp^3$ atoms starts to diminish. Unfortunately, this is an extremely
challenging process to study using highly accurate methods, such as AIMD based on density-functional theory (DFT),
due to their computational cost. Instead,
simulated deposition has been carried out in the past with ``classical'' interatomic potentials such as
Tersoff~\cite{tersoff_1988} and C-EDIP~\cite{marks_2000}. However, classical potentials have systematically failed at reproducing experimentally
observed $sp^3$ fractions~\cite{marks_2005}.
DFT-based generation of a-C has been carried out with varying degree of success using alternative
routes~\cite{marks_1996,mcculloch_2000,marks_2002}. See Ref.~\cite{laurila_2017} for a review of the performance of different
generation methods and potentials.

\begin{figure}[t]
\includegraphics[]{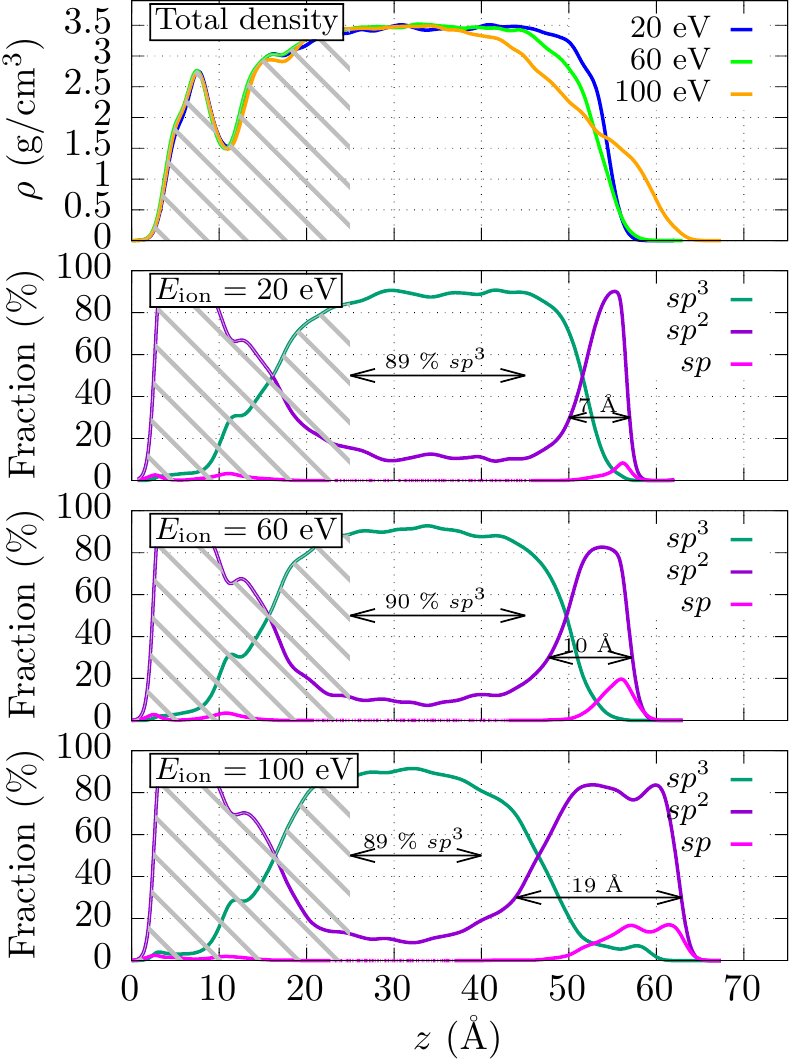}
\caption{Mass density profiles and $sp$, $sp^2$ and $sp^3$ fractions in the bulk of the film,
for the different deposition regimes studied. Atomic coordinations are determined according to a
1.9~{\AA} cutoff radius for nearest neighbors, which corresponds to the first minimum of the
radial distribution function~\cite{caro_2014}.}
\label{01}
\end{figure}

Thus, there is a gap between what would be a close representation of reality and what can be simulated in practice. This gap is due to
the difficulty of modeling realistic processes (large number of atoms, long time scales) and what
can currently be done with accurate, yet computationally-expensive methods, such as DFT-based MD.
Recent advances in computational techniques have given rise to a trend in the physics, chemistry and materials science
communities to apply machine-learning (ML) and data-driven approaches to materials modeling~\cite{khaliullin_2011,sosso_2013}. In the specific
realm of interatomic potentials, a family of general and highly flexible potentials referred to as ``Gaussian approximation
potentials'' (GAPs) has been introduced, which promises to bridge the gap we were referring to earlier~\cite{bartok_2010}.
In this Letter, we use the GAP ML interatomic
potential~\cite{deringer_2017} to study the hitherto unresolved a-C growth mechanism and the physical reasons for the high $sp^3$ concentration
in ta-C films with an unprecedented level of accuracy.

To study the atomistic details of the growth of an a-C film, we explicitly simulated the deposition of C atoms onto
a carbon substrate \textit{one atom at a time}, using MD. A large [111]-oriented diamond substrate, terminated by the stable
$2 \times 1$ surface
reconstruction, was used, containing 3240 atoms in periodic boundary conditions. This corresponds to
initial dimensions of $38 \text{~\AA} \times 38 \text{~\AA}$ in plane and 16~{\AA} of thickness.
The effect of the substrate on the results of the simulation is discussed in the Supplemental Information (SI).
2500 single monoenergetic C atoms with a kinetic energy of 60~eV were dropped from the top of the simulation box onto the diamond substrate, to create
an initial a-C template.
After this, an additional 5500 atoms, each with a kinetic energy corresponding to the different
deposition regimes studied (20~eV, 60~eV and 100~eV), were subsequently deposited, for a total of 8000 impact events per
energy. The equations of motion were integrated using
a time step dynamically adapted to correctly describe the atomic trajectories while maximizing computing efficiency,
ensuring that the largest atomic displacements do not exceed 0.1~{\AA} per time step.
Our main results are obtained with the GAP ML potential trained from local density approximation
(LDA) DFT data~\cite{deringer_2017}.
All MD simulations were carried
out with LAMMPS~\cite{plimpton_1995,ref_lammps}.

The impact of the incident ions \textit{per se} lasts for just a
few fs. However, the kinetic energy of the impacting atom is transferred to the substrate, increasing its temperature.
To ensure that the experimental conditions are met as closely as possible, this extra kinetic energy needs to be removed using
a thermostat, bringing the system back to equilibrium before the next deposition takes place. Equilibrating the system
back to the nominal substrate temperature, 300~K, takes up to 1~ps, depending on the
energy of the incident ion. Equilibration is therefore by far the most computationally expensive part of the simulation.
A more detailed discussion of the dependence on deposition energy (including the low-energy regime),
an in-depth study of elasticity and comparison with Tersoff and C-EDIP results
will be published later in a more technical paper~\cite{caro_2018b}. Video animations of the growth process can be accessed online from
the Zenodo repository~\cite{caro_2017d} and the SI.

In \fig{01} we show the main structural features of the deposited a-C films. The figure shows the in-plane averaged mass density
profile of the films grown at different deposition energies. Very high densities and $sp^3$
fractions are obtained in the interior of the film. The simulated deposition at 60~eV, which is the ion energy at which
$sp^3$ content is expected to peak based on experimental observations~\cite{robertson_2011}, shows $sp^3$ fractions of up to 90\%. Previous
simulations~\cite{mcculloch_2000,marks_2005,caro_2014,laurila_2017},
either based on deposition or alternative methods such as liquid quenching, have systematically failed to
reproduce these high numbers.
The previously reported computational results with the highest $sp^3$ fractions (shy of 85\%) were based on DFT
geometry optimization followed by pressure correction~\cite{caro_2014,laurila_2017}. Explicit deposition simulations
(based on the widely used empirical C-EDIP potential) had not been able to produce a-C structures with $sp^3$ fractions exceeding $\sim$60\%~\cite{marks_2005}.
The 20~eV, 60~eV and 100~eV films from \fig{01} reach mass densities around 3.5 g/cm$^3$, very close to diamond.
Although these densities exceed typical experimental values for ta-C by a few percent, is it indeed possible to grow
``superhard'' ta-C close to the density of diamond under ideal conditions, such as the absence of hydrogen~\cite{schultrich_1998}. Lifshitz
{\etal} showed that ta-C films as dense as 3.5~g/cm$^3$ can be grown consistently over a wide range of deposition energies~\cite{lifshitz_1995},
although we must note that such extremely high-density samples are lacking from most of the literature, where quoted values
are typically below the 3.3~g/cm$^3$ mark.
One also needs to take into consideration that
these ta-C films are under typical compressive stresses equivalent to $\sim 2$~\% change in volume (Table~\ref{06}).

\begin{figure}[t]
\includegraphics[]{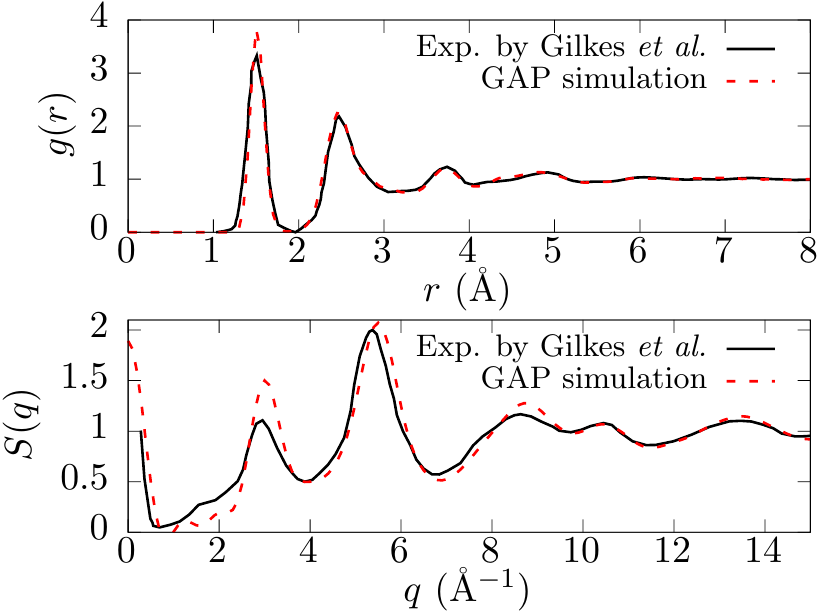}
\caption{Radial distribution function and structure factor in the bulk region of the film, extracted from the 60~eV deposition
simulations, and comparison with experimental data from Gilkes \etal~\cite{gilkes_1995}.}
\label{05}
\end{figure}

The comparison with experimental fingerprints for short and medium range order (\fig{05}) again reveals
excellent agreement and further indicates that GAP provides a correct description of the deposition
physics.
The elastic properties of the films, including stresses built-in during deposition, are summarized in Table~\ref{06}. We note that GAP
has previously been tested to give reliable elastic properties for quenched a-C~\cite{deringer_2017}.
For the present study, we computed the elastic properties of the films in the bulk-like region, that is, the portion of the film where the
$sp^3$ fraction remains constant. Details will be given in a separate paper, which also presents more detailed information on the elastic
properties of the films and their energy dependence~\cite{caro_2018b}.
The data in Table~\ref{06} indeed confirm that ta-C
films are under large compressive stresses, of the order of 10~GPa. Under such compression, this superhard ta-C film is
less compressible than diamond at equilibrium, for which the bulk modulus is $\sim 440$~GPa. The elastic moduli should be significantly
reduced once the strain in the film is released. We observed plastic deformation (bond rearrangement) when attempting film relaxation.
Based on this and on abundant experimental evidence~\cite{robertson_2002}, it is unlikely that highly $sp^3$-rich ta-C can be generated in the
absence of these large compressive stresses. What is more difficult to ascertain is whether compressive stress is required for ta-C
growth or just a consequence of how growth occurs.

\begin{figure}[t]
\includegraphics[]{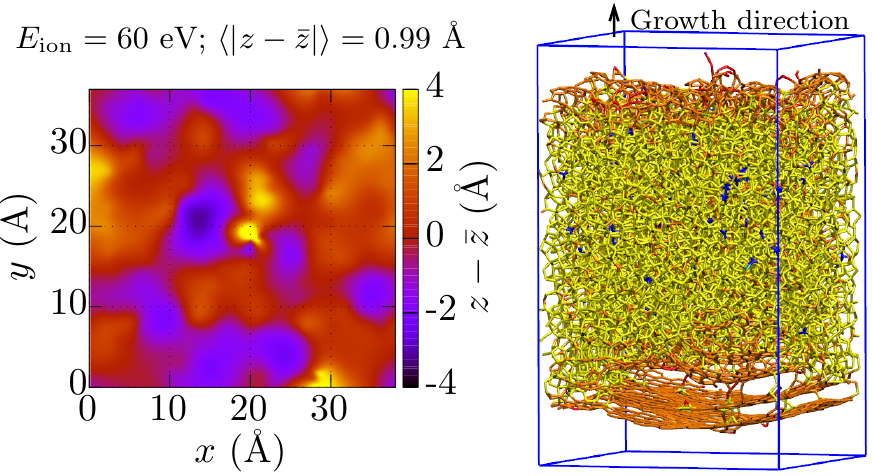}
\caption{Surface roughness and atomic film structure of the 60~eV system, calculated as the mean absolute deviation of
surface height from its average.
Purple, red, orange, yellow and blue atoms represent 1-, 2-, 3-, 4- and 5-fold coordinated C atoms, respectively.
The reason for graphitization of the lower surface and the presence of a few 5-fold coordinated C atoms
are discussed in the SI.}
\label{02}
\end{figure}

\begin{table}[b]
\caption{Elastic properties of the as-grown film (60~eV deposition).
}
\begin{ruledtabular}
\begin{tabular}{l r r}
Quantity & Simulation & Experiment \\
\hline
In-plane stress ($\frac{\sigma_1 + \sigma_2}{2}$) & $-14.4$~GPa & \\
Out-of-plane stress ($\sigma_3$) & $0$~GPa & \\
Stress (isotropic average) & $-9.6$~GPa & $-10$~GPa\footnotemark[1] \\
Equivalent in-plane strain & $-1.4$~\% & \\
Equivalent out-of-plane strain & $0.8$~\% & \\
\hline
Bulk modulus
& $547$~GPa &  $397$~GPa\footnotemark[1] \\
Young's modulus
& $810$~GPa & $760$\footnotemark[1],$850$\footnotemark[2]~GPa\\
\end{tabular}
\footnotetext[1]{Ferrari \etal~\cite{ferrari_1999} for a 3.26~g/cm$^3$ sample. Although the authors report $340$~GPa
as bulk modulus, we note that $397$~GPa is the value which best fits their data when considering the full domain
of elastic moduli compatible with the experimental measurements~\cite{caro_2018b}.}
\footnotetext[2]{Schultrich \etal~\cite{schultrich_1998} for a 3.43~g/cm$^3$ sample.}
\end{ruledtabular}
\label{06}
\end{table}

\begin{figure}[t]
\includegraphics[]{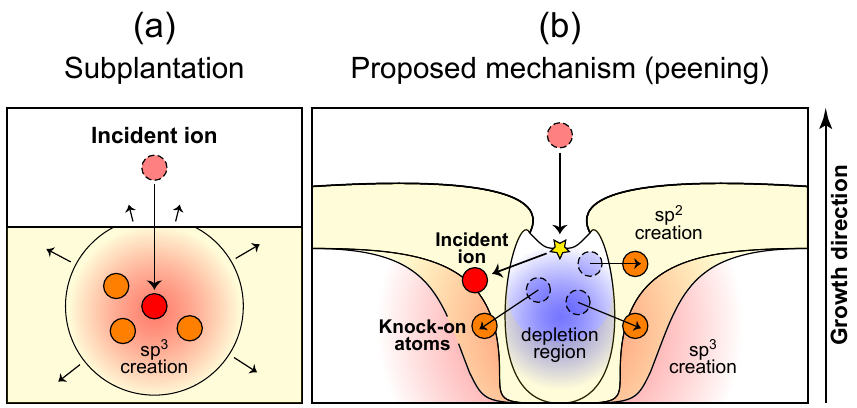}

\includegraphics[]{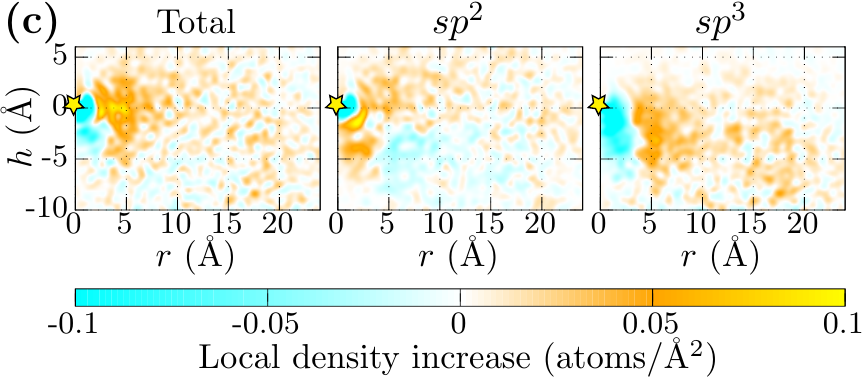}
\caption{(a) Previously accepted growth mechanism in ta-C and (b) growth mechanism
proposed in this Letter. (c) Average increase in local mass density after ion impact (60~eV deposition,
see text for details). The star indicates the impact site.}
\label{04}
\end{figure}

\begin{figure}[t]
\includegraphics[]{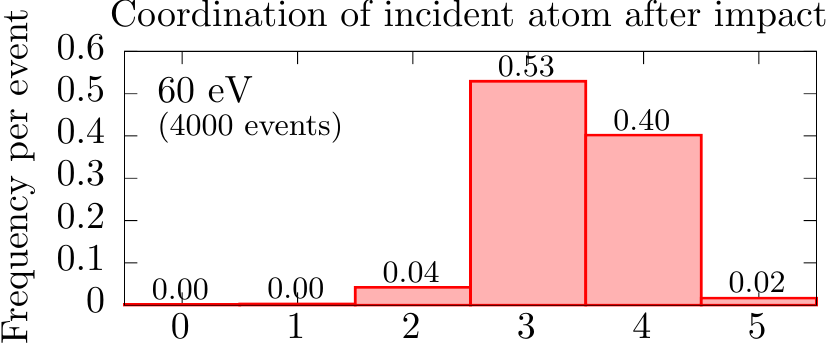}
\includegraphics[width=1\columnwidth]{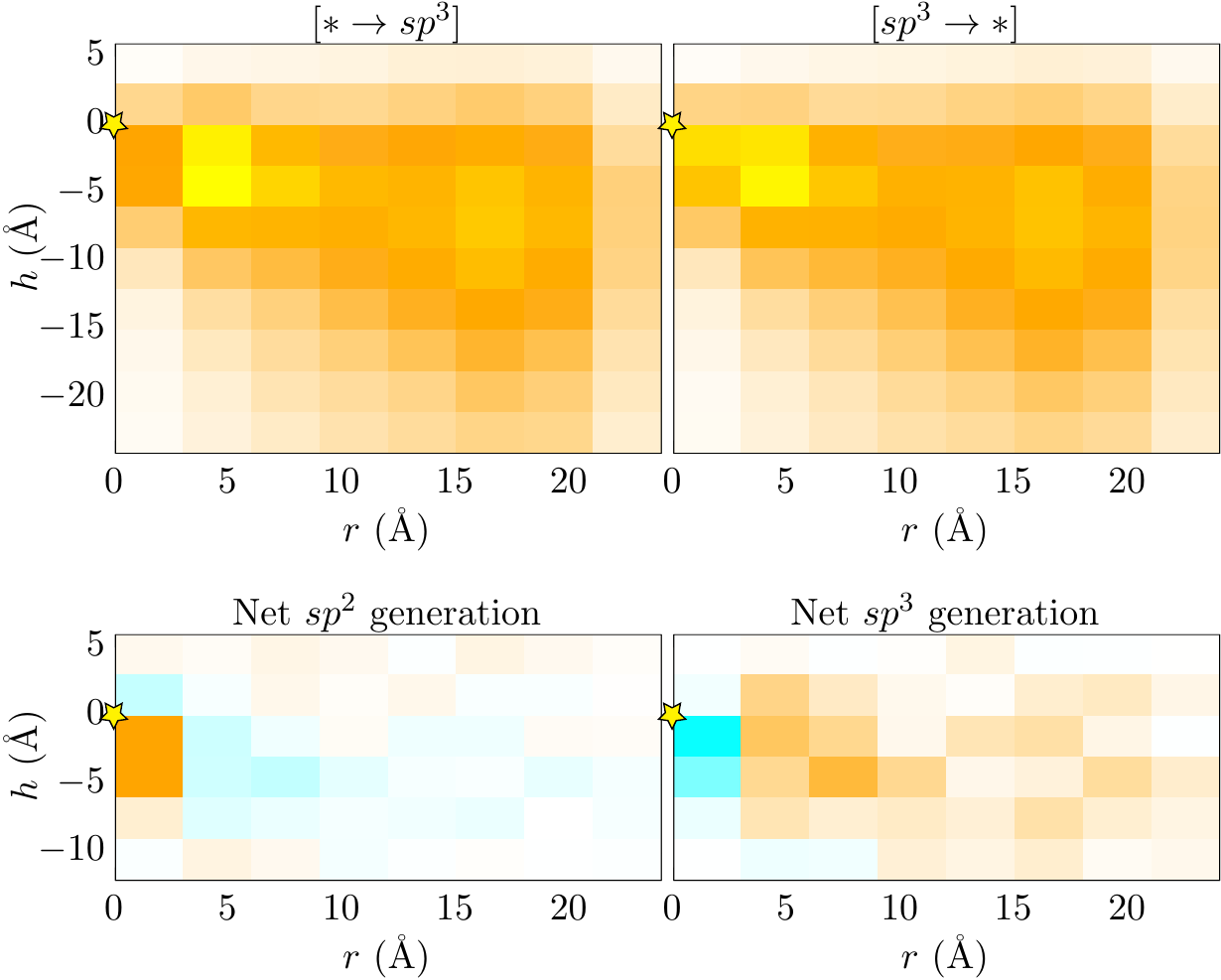}
\caption{(Top) Distribution of coordinations for the incident atom after deposition. (Middle) Average bond
rearrangements that take place for each impact, from and to $sp^3$ coordination, as a function of depth and lateral
distance from impact site. (Bottom) Net generation of $sp^3$ sites and $sp^2$ sites
(both excluding incident atom contribution). Blue, orange and yellow indicate negative, positive and very positive
bond rearrangement, respectively. The star indicates the impact site. An enlarged version of the middle and bottom panels of
this figure, with additional quantitative information, is given in the SI.}
\label{03}
\end{figure}

With regards to surface morphology, \fig{01} already clearly hints toward different features as the deposition energy is varied.
As the ion energy increases, the spatial extent of the $sp^2$-rich region increases too. This can be observed in more detail
in \fig{02}, where we show the final deposited film structure for 60~eV and its topographic surface map. The microscopic surface
roughness for this film is $\sim 1$~{\AA}. We observe that surface
roughness is minimal for the 20~eV film ($\sim 0.7$~{\AA}), and increases for both lower
and higher deposition energies (e.g., $\sim 1.5$~{\AA} and $\sim 1.9$~{\AA} at 5~eV and 100~eV, respectively)~\cite{caro_2018b}.
These results are in qualitative agreement with the detailed experimental study on the
morphology of ta-C surfaces by Davis \etal~\cite{davis_1998}, who measured $\sim 4$~{\AA} and $\sim 10$~{\AA} thick $sp^2$-rich
regions for 35~eV and 100~eV films, respectively. Although Davis' data for surface thickness have large error bars, and the
definition of a ``surface region'' is to some degree arbitrary, we can infer that surface thickness increases experimentally
between 0.1~{\AA/eV} and 0.2~{\AA/eV} within the energy regime relevant to ta-C growth~\cite{davis_1998}. In this context, our estimates of surface
thickness (\fig{01}) also show reasonable quantitative agreement with experiment.
The general conclusion is that the thickness of the surface region grows with deposition energy,
due to the increasing strength of the local thermal spike at the impact site. Impacting atoms induce generation of
$sp^2$-bonded carbon, including \textit{local} transition from $sp^3$ to $sp^2$ coordination.

We now turn our attention
to the microscopic growth mechanism responsible for these high $sp^3$ fractions. The consensus
in the literature is that the ``subplantation'' mechanism is behind this phenomenon~\cite{robertson_2011}. This mechanism
is illustrated in \fig{04} and relates the increase in bonding coordination to the packing of atoms in too small a volume,
as newly arrived atoms are being deposited. The relaxation of the surrounding matrix then explains film growth. However, this
view is in contradiction with the results of our simulations.
While the subplantation mechanism was already challenged by Marks from C-EDIP simulations~\cite{marks_2005},
one of the reasons why an alternative model as already proposed with C-EDIP has not been accepted is the lack of quantitative
agreement with experiment, i.e., the $sp^3$ fractions are too low as predicted by C-EDIP.
In \fig{04} (c) we show the local mass
density difference between the structure before and after impact:
\begin{align}
\Delta \rho (r,h) = 2\pi r \left( g_\text{after}(r,h) - g_\text{before}(r,h) \right),
\label{07}
\end{align}
where $g(r,h)$ is the pair correlation function on the surface of a cylinder of radius $r$ and height $h$ with origin at the
impact site. $\Delta \rho (r, h)$ therefore gives the difference in total atom density integrated on a circumference of radius $r$ around
the impact site at height $h$.
We furthermore resolve this according to $sp^2$ and $sp^3$ components, which are computed with \eq{07} using only the partial local mass densities
corresponding to atoms with 3- and 4-fold coordination, respectively.
This quantity allows us to visualize where atoms are being removed and deposited and where the transition
from $sp^2$ to $sp^3$ is taking place. Orange regions in the color maps
indicate an increase in local density after impact, whereas blue regions denote a decrease in local density. The origin
of the plot, (0,0), corresponds to the impact site, and the maps have been averaged over the last 4000 impacts. Our results
challenge the belief that subplantation explains the high $sp^3$ fractions. The blue region around and
below the impact site on the ``Total'' and ``$sp^3$'' panels shows that atoms are being displaced by the incoming ion.
The orange region circling the impact site in the ``$sp^2$'' panel shows that these atoms, including the incoming ion, are
subsequently deposited preferentially as $sp^2$ atoms.

To further quantify this effect, \fig{03} shows the average changes in atomic coordination
within different regions around the impact site. As mentioned, the impacting atom is preferentially deposited with 3-fold coordination and
there is a net \textit{annihilation} of 4-fold ($sp^3$) sites in the immediate vicinity of the impact site. This is incompatible with
the subplantation mechanism, which would require a majority of impacting atoms to be deposited with 4-fold coordination (see
SI for more quantitative information).
Our data show that each single impact induces coordination changes for roughly 80 atoms, and that
$sp^3$ motifs locally diminish at and around the impact site. However, the dynamical balance between $sp^3$ creation and annihilation builds up
\textit{laterally and away} from the impact region to yield net generation of $sp^3$ carbon as a result.
Figure~\ref{04} (b) shows schematically how the atoms are locally depleted around the impact site and deposited nearby as $sp^2$
carbon. This displacement induces a transformation of the surrounding carbons from $sp^2$ to
$sp^3$, and also the film's growth via vertical displacement of the uppermost layer of C atoms, which are
always predominantly $sp^2$-bonded (and occasionally $sp$). Therefore, our results indicate that the pressure wave generated
by the impacting energetic ions and knock-on atoms is responsible for the generation of $sp^3$-rich a-C films.
This process is beneficial at the studied 20~eV, 60~eV and 100~eV deposition energies, but it does not occur at lower energies~\cite{caro_2018b}.
As the deposition energy increases, the incoming ions carry enough kinetic energy to start damaging the surface, which leads to
the creation of a thicker and more disordered $sp^2$ surface region (Figs.~\ref{01} and \ref{02}), in agreement with experiment~\cite{davis_1998}.

To summarize, this is the first computational
study to report deposited a-C structures with a degree of $sp^3$ hybridization in quantitative agreement with experiment.
Most importantly, the excellent agreement that we obtain with relevant experiments
gives us confidence that our simulation is reproducing the microscopic physical processes correctly.
In turn, this gives us confidence that we provide a fully atomistic account
of the growth mechanism and high $sp^3$ contents in ta-C. The growth mechanism clearly supported by our results is
peening; the previously proposed subplantation mechanism cannot be substantiated in view of our data. The use of a machine-learned
interatomic potential trained from \textit{ab initio} data has allowed us to achieve a level of description for this complex problem
that has previously been out of reach. We believe these results also highlight the role that machine learning will play in the field of
materials modeling and molecular dynamics in the years to come.

This research was financially supported by the Academy of Finland through grants 310574 and
285526. Computational resources were provided by CSC -- IT Center for Science, Finland, though projects 2000634 and 2000300.
V.~L.~D. gratefully acknowledges a fellowship from the Alexander von Humboldt Foundation, a Leverhulme Early Career Fellowship,
and support from the Isaac Newton Trust.

\end{document}